\documentclass[10pt, notitlepage, twocolumn, superscriptaddress, nofootinbib, floatfix]{revtex4-1}

\usepackage{amsmath}
\usepackage{mathrsfs}
\usepackage{multirow}
\usepackage{amssymb}
\usepackage{mathtools}
\usepackage{hyperref}
\usepackage[usenames,dvipsnames]{xcolor}
\hypersetup{colorlinks=true, citecolor=Purple, linkcolor=Purple,
urlcolor=Purple}
\usepackage{soul}
\usepackage{graphicx}
\graphicspath{{./Figures/}}
\usepackage{orcidlink}

\newcommand{\beq}{\begin{equation}}
\newcommand{\eeq}{\end{equation}}
\newcommand{\ba}{\begin{eqnarray}}
\newcommand{\ea}{\end{eqnarray}}

\definecolor{tclr}{RGB}{148,0,211}

\begin{document}

\allowdisplaybreaks

\title{Greybody factors in scalar-tensor gravity and beyond}

\author{Georgios~Antoniou~\orcidlink{0000-0002-5974-320X}}
\email{georgios.antoniou@roma1.infn.it}
\affiliation{\href{https://www.phys.uniroma1.it/en}{Dipartimento di Fisica, ``Sapienza'' Universit\'a di Roma}, P.A. Moro 5, 00185, Roma, Italy}
\affiliation{\href{https://www.roma1.infn.it/en/home.html}{Sezione INFN Roma1}, P.A. Moro 5, 00185, Roma, Italy}

\author{Thomas D.~Pappas~\orcidlink{0000-0003-2186-357X}}
\email{thomas.pappas@physics.slu.cz}
\affiliation{\href{https://www.slu.cz/phys/en/}{Research Centre for Theoretical Physics and Astrophysics,\\ Institute of Physics, Silesian University in Opava},\\ Bezručovo náměstí 13, CZ-746 01 Opava, Czech Republic}

\author{Panagiota~Kanti~\orcidlink{0000-0002-3018-5558}\,}
\email{pkanti@uoi.gr}
\affiliation{\href{https://theory.physics.uoi.gr/}{Division of Theoretical Physics, Department of Physics, University of Ioannina}, GR 45500, Ioannina, Greece}

\begin{abstract}
In the framework of the beyond Horndeski action, we consider three subtheories that support scalarised black-hole solutions, and look for modified characteristics compared to GR. We first study the propagation of massless scalar and vector test fields in the fixed background of an analytical spherically symmetric black hole derived in the context of a parity-symmetric beyond Horndeski theory, and show that the profiles of the effective gravitational potentials, greybody factors, absorption cross sections and quasinormal frequencies exhibit distinct modifications as we move away from the GR limit. We then turn our attention to the perturbations of the gravitational field itself and adopt a full-theory analysis that takes into account the backreaction of the scalar field on the metric. Employing as background solutions scalarized black holes arising in the shift-symmetric Horndeski theory and in the quadratic-quartic-scalar-Gauss-Bonnet theory, we compute the greybody factors and quasinormal modes of the axial sector. In both theories, in direct correspondence to the form of the gravitational potential, which features multiple extremal points, we find modified (suppressed or nonmonotonic) greybody curves and altered quasinormal frequencies (smaller oscillating frequencies and larger damping times), especially as the hair-sourcing parameter increases.

\end{abstract}

\maketitle

%===============================================
\section{Introduction}
\label{Sec:Introduction}
%===============================================

As we have long ago come to realize, the most mysterious objects in our universe, black holes (BHs), keep their secrets well and deeply hidden. According to general relativity (GR), no particle, signal, or piece of information can escape from within its event horizon, and only a small number of distinct quantities can characterize a black hole: its mass, electromagnetic charge, and angular momentum~\cite{Bekenstein:1971hc, Price:1971fb, Teitelboim:1972ps}. The possibility of the emergence of scalar hair was also debated in the early days of scalar-tensor theories of gravity, leading eventually to the corresponding no-hair theorems~\cite{Bekenstein:1974sf, Bekenstein:1975ts, Bekenstein:1995un}. The revival of Horndeski theory~\cite{Horndeski:1974wa}, as the most general framework of a scalar-tensor theory leading to field equations with up to second-order derivatives, caused the formulation of new no-scalar-hair theorems~\cite{Sotiriou:2011dz, Hui:2012qt}.

However, all scalar no-hair theorems were proved to be rather short lived. Solutions with scalar hair, although secondary, were found quite early on (see~\cite{Callan:1988hs, Campbell:1990ai, Mignemi:1992pm, Kanti:1995cp, Kanti:1995vq, Torii:1996yi, Guo:2008hf, Kleihaus:2011tg, Pani:2011gy} for some indicative early works and the reviews~\cite{Charmousis:2008kc, Sotiriou:2014yhm, Herdeiro:2015waa} for a more exhaustive list). In the new era for scalar-hairy black holes that was launched with the revival of Horndeski theory, more solutions soon followed~\cite{Babichev:2013cya, Sotiriou:2013qea, Herdeiro:2014goa, Sotiriou:2014pfa, Babichev:2016rlq, Antoniou:2017acq, Doneva:2017bvd, Silva:2017uqg, Antoniou:2017hxj}, which in turn paved the way for a plethora of additional, similar solutions to be found. 

The Horndeski and beyond Horndeski actions~\cite{Gleyzes:2014dya, Langlois:2015cwa, Crisostomi:2016czh, Crisostomi:2016tcp, Kobayashi:2019hrl} provide a rather general and flexible mathematical framework, parametrized by a number of coupling functions. Upon carefully choosing the form of these coupling functions, physically interesting black-hole solutions with attractive characteristics, often modified compared to traditional GR, emerge. These solutions along with any observable quantities associated with them may provide the way to explore the fundamental theory of gravity and discover new physics in the strong gravity regime.

Greybody factors (GB) are physical quantities that characterize every scattering process occurring in the gravitational background of a black hole. They carry information about the type of particle that undergoes the scattering process but also about the form of the line-element--and thus the conserved charges and fundamental parameters--of the spacetime itself. They have been used extensively in the literature to study aspects not only of GR but also of generalized theories of gravity, either string-inspired or higher-dimensional ones~\cite{Hawking:1975vcx, Carter:1974yx, Page:1976df, Unruh:1976db, Sanchez:1977si, MacGibbon:1990zk, Das:1996we, Maldacena:1996ix, Gubser:1997yh, Cvetic:1997uw, Kanti:2004nr}. 

Quasinormal modes (QNMs) are also a valuable tool for the study of black holes as they characterize the way a propagating field or the spacetime itself responds to a perturbation~\cite{Nollert:1999ji,Kokkotas:1999bd,Konoplya:2011qq,Konoplya:2019hlu}. They describe both the resonant frequencies of the oscillations produced as a result of the perturbation as well as the damping times that these oscillations take to die out. QNMs are therefore directly related to the dynamics and stability of the associated fields.

Both greybody factors and quasinormal frequencies may be derived by solving the same field equations but under different boundary conditions as they correspond to different settings. However, an underlying connection exists as both quantities draw their characteristics from the form of the effective gravitational potential which is common in both processes. Recently, a direct correspondence between greybody factors and QNMs was initially conjectured in~\cite{Kyutoku:2022gbr,Oshita:2023cjz,Pedrotti:2025upg} and later formulated more precisely in~\cite{Konoplya:2024lir} (see also~\cite{Bolokhov:2024otn, Malik:2024cgb, Malik:2024wvs, Rosato:2024arw, Oshita:2024fzf, Rosato:2025byu}). According to this, for the class of effective potentials for which the Wentzel–Kramers–Brillouin (WKB) expansion is valid~\cite{Konoplya:2019hlu}, the greybody factors can be well approximated in terms of analytic formulas that involve the QNMs. The correspondence was initially formulated in the case of a spherically symmetric black hole but also extends to rotating black holes~\cite{Konoplya:2024vuj}. Although there might not be such a direct correspondence in the general case where the gravitational potential deviates from a well-behaved, single-peak form that is compatible with the WKB method, an interconnection or a common behavior is expected to be again observed.

In the present work, we will focus on subclasses of Horndeski and beyond Horndeski theory, and perform a thorough investigation of both the greybody factors and spectrum of quasinormal frequencies. In the first part of this manuscript, we will adopt the more conventional approach for the derivation of greybody factors, and solve the equations of massless scalar and vector test fields propagating in the fixed background of a spherically symmetric scalar-hairy black hole emerging in the context of the parity-symmetric beyond Horndeski theory~\cite{Bakopoulos:2022csr} (see also~\cite{Konoplya:2020cbv, Grain:2005my,Lima:2020seq,Dubinsky:2024vbn} for similar analyses in different Horndeski or hairy solutions). We will study the form of the gravitational potentials that the two types of propagating fields feel, and derive the greybody factors via the transmission probabilities in the specific gravitational background. We will also compute in each case the dimensionful quantity of the total absorption cross-section, and study both its low- and high-energy limits. Changing appropriately the boundary conditions, the QNM spectrum will also be derived.~The dependence of all the aforementioned quantities on the values of the fundamental parameters of the theory will be studied, and modifications compared to GR will be sought.

In the second part of our manuscript, we will address the perturbations of the gravitational field itself.~We will adopt a full-theory approach, and consider also the coupled perturbations of the scalar field of the theory. We will consider two different subclasses of Horndeski theory, the shift-symmetric and the quadratic-quartic-scalar-Gauss-Bonnet theory. Here, we will work in the linear approximation, and focus on the axial sector where the perturbations of the scalar and gravitational fields get decoupled. We will derive and study the gravitational potentials in each case, and for the sake of comparison, we will present also the corresponding results for a test spin-2 field. Applying the appropriate boundary conditions, we will determine both the greybody factors and quasinormal frequencies, and study their behavior for a number of background black-hole solutions along the complete existence line spanned by the hair-sourcing parameter. To our knowledge greybody factors have not been calculated before in the context of modified gravities with additional scalars in the nondecoupling limit. We therefore hope that our analysis will pave the way for similar analyses in other frameworks, while simultaneously complementing and expanding the relevant studies that have appeared in the literature~\cite{Tattersall:2018nve, Yang:2023lcm, Antoniou:2024gdf, Antoniou:2024hlf, Yang:2024cjf, Charmousis:2025xug}. 

The outline of our paper is as follows:~we start in Sec.~\ref{Sec:II} with the presentation of the general theoretical framework and form of spacetime background. In Sec.~\ref{Sec:III}, we perform the test-field analysis for scalar and vector particles propagating in the background of a spherically symmetric scalarized black hole of beyond Horndeski theory. In Sec.~\ref{Sec:IV}, we proceed to perform the full-field analysis of the perturbations of the gravitational and scalar field, and study the axial sector employing as background solutions two spherically symmetric black holes emerging in the context of Horndeski theory. We finish with our conclusions in Sec.~\ref{Sec:Conclusions}.

%===============================================
\section{Theoretical Framework}
\label{Sec:II}
%===============================================

As a starting point of our analysis, in this section, we present the action functional of the general class of theories in which we will work, and the form of the spacetime background of the solutions whose properties we will study.

%------------------------------------%
%------------------------------------%
\subsection{The theory }
%------------------------------------%
%------------------------------------%
The general theoretical framework of our analysis will be that of beyond Horndeski theory defined by the action
\begin{equation}
    S=S_\text{H}+S_\text{bH},
\label{eq:framework}
\end{equation}
where
\begin{align}
S_\text{H} = &\int d^4 x \sqrt{-g} \,\left(\mathcal{L}_2+\mathcal{L}_3+\mathcal{L}_4+\mathcal{L}_5\right)\,,\\
S_\text{bH} = & \int d^4 x \sqrt{-g} \,\left(\mathcal{L}^{\rm bH}_4+\mathcal{L}^{\rm bH}_5\right)\,,
\end{align}
describe the Horndeski and beyond Horndeski terms, respectively~\cite{Horndeski:1974wa, Gleyzes:2014dya, Langlois:2015cwa, Crisostomi:2016czh, Crisostomi:2016tcp, Kobayashi:2019hrl}. The Lagrangian terms ${\cal L}_i$ of the Horndeski theory are given by the following expressions
\begin{align}
\mathcal{L}_2 &= \;G_2(X) ,
\label{eq:L2fr}
\\
\mathcal{L}_3 &= \;-G_3(X) \,\Box \phi ,
\label{eq:L3fr}
\\
\mathcal{L}_4 &= \;G_4(X) R + G_{4X} \left[ (\Box \phi)^2 -\nabla_\mu\nabla_\nu\phi \,\nabla^\mu\nabla^\nu\phi\right] ,
\label{eq:L4fr}
\\
\begin{split}
\mathcal{L}_5 &= \;G_5(X) G_{\mu\nu}\nabla^\mu \nabla^\nu \phi - \frac{1}{6}\, G_{5X} \big[ (\Box \phi)^3\\
&- 3\,\Box \phi\, (\nabla_\mu\nabla_\nu\phi)^2+ 2\,\nabla_\mu\nabla_\nu\phi\, \nabla^\nu\nabla^\rho\phi\, \nabla_\rho\nabla^\mu\phi \big],
\label{eq:L5fr}
\end{split}
\end{align}
where $\phi$ is a scalar degree of freedom, $X \equiv -\partial_\mu \phi\,\partial^\mu \phi/2$ stands for its kinetic term and $G_i(X)$, with $i=2,3,4,5$, are arbitrary functions of $X$. The beyond Horndeski theory follows via the addition to $S_H$ of the two higher-order terms
\begin{align}
\mathcal{L}^{\rm bH}_4&=F_4(X)\varepsilon^{\mu\nu\rho\sigma}\,\varepsilon^{\alpha\beta\gamma}_{\,\,\,\,\,\,\,\,\,\,\,\sigma}\,\partial_\mu\phi\,\partial_\alpha\phi\,\nabla_\nu\partial_\beta\phi\,\nabla_\rho\partial_\gamma\phi,\\[3mm]
\mathcal{L}^{\rm bH}_5&=F_5(X)\varepsilon^{\mu\nu\rho\sigma}\,\varepsilon^{\alpha\beta\gamma\delta}\,\partial_\mu\phi\,\partial_\alpha\phi\,\nabla_\nu\partial_\beta\phi\,\nabla_\rho\partial_\gamma\phi\,\nabla_\sigma\partial_\delta\phi,
\end{align}
which are parametrized by two additional coupling functions $F_4$ and $F_5$. These two functions are not independent but related via the following relation~\cite{Crisostomi:2016tcp}
\begin{equation}
    X G_{5X} F_4= 3 F_5 (G_4-2X G_{4X})\,,
\end{equation}
in order to evade the appearance of a ghost degree of freedom.

%------------------------------------%
%------------------------------------%
\subsection{The spacetime background }
%------------------------------------%
%------------------------------------%

In this work, we will focus on static, spherically symmetric black-hole solutions with the spacetime background around them given by the line element
\begin{equation}
    d s^2=-A(r)\,d t^{2} +\frac{dr^2}{B(r)}+ r^{2} \left( d \theta^{2}+\sin^{2}\theta\,d\varphi^{2} \right).
\label{eq:metric}
\end{equation}

The exact form of the two metric functions $A(r)$ and $B(r)$ is determined via Einstein's field equations coupled to the scalar field equation, as these follow from the action functional of the theory.~The above line element serves as the fixed spacetime background in which a test field may propagate or as the background gravitational solution when perturbations of the full solutions are considered. 

Depending on the exact form of the theory, the metric functions and form of the scalar field $\phi(r)$ may be determined either analytically or via numerical integration of the field equations. In either case, their series expansions near the black hole horizon and at asymptotic infinity are useful to know.
The relevant expansions near $r_h$ take the form
\begin{align}
    A(r)=&\sum_{n=1}a^{(n)}(r-r_h)^n, \label{eq:expA}\\
    B(r)=&\sum_{n=1}b^{(n)}(r-r_h)^n , \label{eq:expB}\\
    \phi(r)=&\;\phi_h+\sum_{n=1}\phi^{(n)}(r-r_h)^n ,\label{eq:phi}
\end{align}
while at infinity we write 
\begin{align}
    A(r)=&1-\frac{2M}{r}+\sum_{n=1}\frac{\tilde{a}^{(n)}}{r^n},\label{eq:expAi}\\
    B(r)=&1-\frac{2M}{r}+\sum_{n=1}\frac{\tilde{b}^{(n)}}{r^n}, \label{eq:expBi}\\
    \phi_0(r)=&\frac{Q}{r}+\sum_{n=2}\frac{\tilde{\phi}^{(n)}}{r^n},\label{eq:phii}
\end{align}
where $M$ and $Q$ are the Arnowitt–Deser–Misner (ADM) mass and the scalar charge, respectively.~The value of the scalar field at the horizon $\phi_h$ is determined via a shooting method so that the asymptotic value of the scalar field vanishes. We also fix $a^{(0)}$ so that we retrieve the Minkowskian metric at infinity.
The remaining coefficients are determined by substituting the expansions in the field equations and solving them order by order.

%===============================================
\section{Test-Field Analysis}
\label{Sec:III}
%===============================================

The simplest approach in which one may derive both greybody factors and QNM frequencies is to consider massless test scalar and electromagnetic fields propagating in a fixed gravitational background such as the one given in Eq.~\eqref{eq:metric}. The corresponding equations of motion for these fields are
\begin{align}
    \Box\Phi=0\, ,\label{eq:test_scalar}\\
    \nabla^\nu F_{\mu\nu}=0\,. \label{eq:test_EM}
\end{align}

We will assume a factorized ansatz for the propagating fields in the form of partial waves according to the following expression
\begin{equation}
    {\tilde \Psi}(t,r,\theta,\varphi)= \int d{\tilde \omega} \,\frac{\Psi(r)}{r}\,Y_{\ell}^{m}(\theta,\varphi)\,e^{-i \tilde\omega t}\,,\label{eq:test_decomposition}
\end{equation}
where $\tilde \Psi\equiv(\Phi,\mathcal{A})$, $Y_{\ell}^{m}(\theta,\varphi)$ are the spherical harmonics and $\tilde \omega$ is the frequency of the partial wave. In that case,
the test-field equations~\eqref{eq:test_scalar}-\eqref{eq:test_EM} result in the following ``master'' equation for the radial part of the propagating fields 
\begin{equation}
    \Psi''+\frac{(A B)'}{2AB}\,\Psi'+\bigg[\frac{\tilde\omega^2}{A B}-\frac{\ell  (\ell +1)}{B r^2}-(1-s^2)\frac{(A B)'}{2 A B r}\bigg] \Psi =0\,,
\end{equation}
where $s=0,1$ for the scalar and electromagnetic field, respectively. The above equation can be recast into a Schr\"odinger-type equation of the form
\begin{equation}
    \frac{d^2\Psi}{dr_*^2}+\left[\tilde\omega^2-\frac{\ell(\ell+1)A}{r^2}-(1-s^2)\frac{(A B)'}{2r}\right] \Psi=0 \label{eq:Schrodinger}
\end{equation}
when the tortoise coordinate, defined as $dr_*/dr=(A B)^{-1/2}$ is employed (see e.g.~\cite{Konoplya:2006rv}).

The greybody factors are connected to the transmission coefficients associated with the scattering process that an incoming test field undergoes in the vicinity of the black hole as a result of the potential gravitational barrier. Therefore, the appropriate boundary conditions read 
\begin{align}
    \Psi=& \,e^{-i \Omega r_*} + R_{\ell}(\Omega)\,e^{i \Omega r_*}\,, \quad r_* \rightarrow +\infty\,, \nonumber \\
   \Psi=& \,A_{\ell}(\Omega)\,e^{-i \Omega r_*}\,, \quad \qquad \quad \,\,r_* \rightarrow -\infty\,,\label{eq:transmission}
\end{align}
where $R_{\ell} (\Omega)$ and $A_{\ell}(\Omega)$ are the reflection and absorption (or transmission) coefficients, respectively, having normalized the amplitude of the incoming wave at infinity to unity. We note that we have denoted the real-valued frequency of the incoming propagating field with $\Omega$. The greybody factor is then defined as
\begin{equation}
    \Gamma_\ell(\Omega) \equiv \,|A_{\ell} (\Omega)|^2 = 1- |R_{\ell}(\Omega)|^2\,.
\end{equation}
Solving Eq.~\eqref{eq:Schrodinger} with the boundary conditions~\eqref{eq:transmission} allows us to determine the greybody factor $\Gamma_{\ell}(\Omega)$ of the $\ell$ partial wave with frequency $\Omega$.

Equation~\eqref{eq:Schrodinger} can also be used to study the quasinormal modes that are associated with perturbations of the gravitational background. These perturbations may be induced directly on the form of the spacetime but they can also be triggered by the propagation of test fields in the vicinity of the black hole, as considered above. The quasinormal modes have characteristic complex frequencies, which henceforth will be denoted by $\omega$, and satisfy boundary conditions appropriate for purely outgoing waves at infinity and purely ingoing waves at the black-hole horizon. The latter therefore read
\begin{align}
    \Psi=& \,e^{i \omega r_*}\,, \quad \,\,\,r_* \rightarrow +\infty\,, \nonumber \\
   \Psi=& \,e^{-i \omega r_*}\,, \quad r_* \rightarrow -\infty\,.\label{eq:QNM}
\end{align}
When the effective potential of the Schr\"odinger-type equation~\eqref{eq:Schrodinger} has the form of a single-peak barrier, the WKB method may be used to determine the frequencies of the dominant quasinormal modes. The first-order WKB formula corresponds to the eikonal approximation and becomes exact in the limit $\ell \rightarrow \infty$. In this approximation, the frequencies of the dominant modes are given by 
\begin{equation}
    \omega = \ell \sqrt{U_{0}} - \frac{i}{2}\,\sqrt{\frac{-U''_{0}}{2U_{0}}} + {\cal O}(\ell^{-1})\,,
    \label{eq:fundamental}
\end{equation}
where $U_0$ is the first-order term in the expansion of the effective potential in terms of $\ell$ around its peak, namely
\begin{equation}
    V(r_*)  \simeq \ell^2 U_0(r_*) + \ell U_1(r_*) + U_2(r_*) + \ell^{-1} U_3(r_*) + ...
\end{equation}
while the primes in $U_0$ in Eq.~\eqref{eq:fundamental} denote the second derivative applied on $U_0$. The quasinormal frequencies provide valuable information on the stability of the background solution with modes having $\textrm{Im}(\omega)>0$ denoting unstable configurations.

\begin{table} 
    \centering
    \label{table:scalarization_thresholds}
    \begin{tabular}{| l | c | c | c |}
    \hline
    \multicolumn{4}{|c|}{GR fundamental mode $M\omega_{0\ell}$} \\[1mm]
    \hline\hline
    Type & $\ell=0$ & $\ell=1$ & $\ell=2$\hspace{0mm} \\[1mm]
    \hline
    \text{scalar} & $0.1104 - 0.1049 i$ & $0.2929-0.0977 i$ & $0.3737-0.0890i$ \\
    \text{vector} & - & $0.2483-0.0925i$ & $0.4576-0.0950i$ \\
    \text{tensor} & - & - & $0.3737-0.0889i$ \\
    \hline
    \end{tabular}
\caption{QNMs for $s=0,1,2$ (scalar, vector, tensor) perturbations on a Schwarzschild background.}
\label{tab:GR_QNMs}
\end{table}

In Table~\ref{tab:GR_QNMs}, we present a summary of the quasinormal frequencies for the fundamental modes of scalar and vector fields propagating on a fixed Schwarzschild background. For completeness, we list also here the corresponding QNMs of the gravitational field itself which will be relevant in the analysis of Sec.~\ref{Sec:IV}. 
We will be using these values as the appropriate GR limits while implementing shooting methods in the following sections.

%------------------------------------------------%
\subsection{Spherical solutions beyond Horndeski theory}
%------------------------------------------------%

Within the extended framework~\eqref{eq:framework} of beyond Horndeski theory, exact analytical solutions describing spherically symmetric black holes have been obtained in~\cite{Bakopoulos:2022csr}. In particular, the parity symmetric theory with $G_3=G_5=F_5=0$, and
\begin{align}
    G_2=& \,-\epsilon \mu\,X^2\,, \nonumber \\[1mm]
   G_4=& \,-\frac{\delta\mu}{2}\,X^2 + \frac{\beta -\delta \zeta}{2}\,X +1\,, \label{eq:G_BeyHorn} \\
   F_4=& \,\frac{\delta \zeta -\beta}{8X} + \frac{3 \delta \mu}{8}\,, \nonumber
\end{align}
was considered along with a static, spherically symmetric scalar field $\phi=\phi(r)$. Note that $G_4$ contains a constant term equal to unity; therefore, this theory accommodates the Einstein-Hilbert term for gravity. In the above expressions, 
$(\beta, \delta, \epsilon, \zeta, \mu)$ are coupling parameters of the theory\footnote{For some interesting generalizations of this theory, which lead to alternative black-hole solutions that could also be employed in a similar analysis, see \cite{Bakopoulos:2023fmv, Baake:2023zsq}.}. 

The theory~\eqref{eq:G_BeyHorn} allows for asymptotically flat black-hole solutions for which the metric functions have the following functional form
\beq
A(r)=B(r)=1+\frac{p_2\,\arctan{\left(p_1\,r\right)}}{p_1\,r}-\frac{2 M}{r}\,,
\label{eq:BCKL_metric}
\eeq
where $p_1 \equiv \sqrt{\epsilon/\delta} > 0$ and has dimensions of (length)$^{-1}$, $M$ is an integration constant with dimensions of (length) and $p_2 \equiv (\beta -\delta \zeta)^2/(8 \delta \mu)$ is dimensionless. The asymptotic expansion of Eq.~\eqref{eq:BCKL_metric} at spatial infinity leads to
\beq
A(r) \simeq 1-\frac{2M_{\rm ADM}}{r}+\frac{Q^2}{r^2}\,,
\label{eq:BCKL_inf}
\eeq
where the ADM mass and tidal charge are given by
\beq
M_{\rm ADM}=M-\frac{p_2 \pi}{4 p_1}\,,\quad Q^2=-\frac{p_2}{p_1^2}\,.
\eeq
Therefore, for $p_2<0$, the solution exhibits a robust Reissner-Nordstr\"om asymptotic limit with $Q^2>0$~\cite{Bakopoulos:2022csr}. In the limit $p_2 \to 0$, Eq.~\eqref{eq:BCKL_metric} reduces to the Schwarzschild BH of mass $M$, and the first derivative of the scalar field, being proportional to $p_2$~\cite{Bakopoulos:2022csr}, vanishes everywhere, thus, recovering GR.

%%%%%%%%%%%%%%%%%%%%%%%%%%%%%%%%%%%%%%%%%%%%%%%%%%%%%%%%%%%%%%%%%%%%
\begin{figure*}[t!]
\centering
\includegraphics[width=0.49\textwidth]{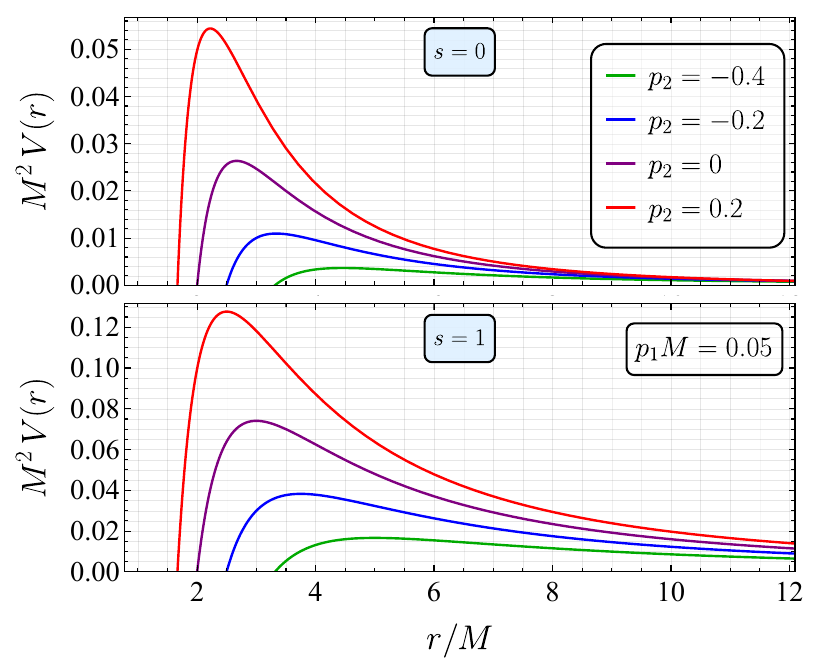}
\includegraphics[width=0.49\textwidth]{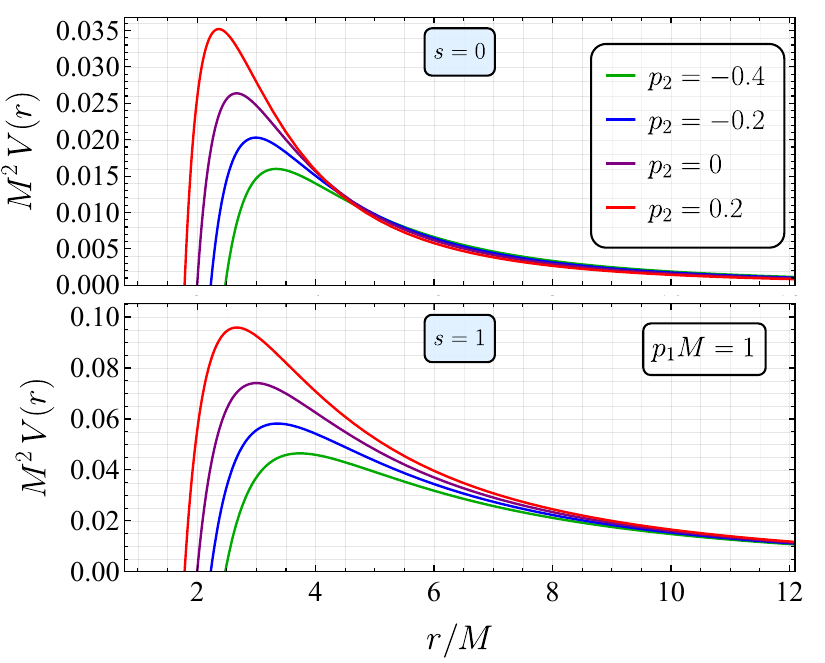}
\caption{Dominant mode $(l=s)$ effective potentials for scalar $(s=0)$ and electromagnetic (EM) $(s=1)$ test-field perturbations in the background~\eqref{eq:BCKL_metric}, left panel for~$p_1\,M=0.05$ and right panel for~$p_1\,M=1$.}
\label{fig:Veffs_BCKL}
\end{figure*}
%%%%%%%%%%%%%%%%%%%%%%%%%%%%%%%%

The event-horizon radius for the black hole described by Eq.~\eqref{eq:BCKL_metric} follows by solving the equation $A(r)=0$, however, this cannot be performed analytically in general. Nevertheless, in the limit $p_1\,r\ll1$, the metric function~\eqref{eq:BCKL_metric} can be approximated by
\beq
A(r)\simeq 1-\frac{2M}{r}+p_2-\frac{p_2\,p_1^2}{3}r^2\,,
\label{eq:BCKL_p1_approx}
\eeq
and the equation $A(r)=0$ can be solved analytically~\cite{Holmes_2002,Nickalls_2006} to give 
\beq
r_h\simeq2\sqrt{-\frac{1+p_2}{p_1^2\,p_2}}\sinh{\left[\frac{1}{3}\sinh^{-1}{\left(\frac{3 M}{1+p_2}\sqrt{-\frac{p_1^2\,p_2}{1+p_2}}\right)} \right]}\,,
\label{eq:rh_p2_neg}
\eeq
for the physically acceptable case with $p_2<0$ and when $9\,M^2p_1^2\,p_2-\left(1+p_2 \right)^3<0$. In order to get a feeling for the type of modifications that arise compared to the GR case in the near-horizon regime, working in the limit of small $p_2$, Eq.~\eqref{eq:rh_p2_neg} reduces to
\beq
r_h=2M+\left(-2M+\frac{8 M^3p_1^2}{3} \right)p_2+\mathcal{O}\left(p_2^2\right)\,.
\label{eq:rh_Schw_lim}
\eeq
We observe that, when $p_1 M \leqslant\sqrt{3}/2$, negative values of $p_2$ result in an increased radius for the event horizon with respect to~the Schwarzschild case. As a result, these black holes are more sparse compared to their GR analogs with the same mass. This behavior remains accurate, with the error being smaller than $1\%$, for all solutions characterized by values of $p_1 M \leq 0.1$ and $p_2\geqslant-0.6$

Employing the metric functions~\eqref{eq:BCKL_metric} in the Schr\"odinger-type equation~\eqref{eq:Schrodinger} of the test scalar and vector fields, we may determine the form of the effective potentials that these fields ``feel'' as they propagate in the corresponding black-hole background. The effective potentials for the scalar $(s=0)$ and vector $(s=1)$ test fields are shown in ~Fig.~\ref{fig:Veffs_BCKL}, for the dominant modes with multipole numbers $l=s$. In both cases, the effective potentials have the form of a single-peak gravitational barrier. In ~Fig.~\ref{fig:Veffs_BCKL}, we depict the effective potentials for two indicative values of $p_1 M$ and for a number of values of $p_2$--in the latter case, for completeness, we also show the corresponding profile for $p_2=0$ (the Schwarzschild case) and $p_2>0$. The height of the barrier increases, for both types of fields, as $p_2$ takes on values that gradually change from negative to zero and then to positive ones. The impact of the parameter $p_1 M$ is a more subtle one and depends on the sign of $p_2$. Focusing on the physically interesting case with $p_2<0$, we observe that the height of the barrier increases as $p_1 M$ takes on larger values. This can be understood by the observation that, for $p_1\,M\gg1$, Eq.~\eqref{eq:BCKL_metric} approaches the Schwarzschild metric and as such, the deviations from the Schwarzschild limit induced by the nonzero value of $p_2$ become less pronounced. We observe that in general the barrier is higher for vector fields compared to the one for scalars. Finally, we note that, for higher multipole numbers, the impact of $p_2$ and $p_1$ is qualitatively the same as for the dominant modes, however, the height of the barrier increases with $\ell$ as expected. 

%-------------------
\subsubsection{Greybody factors and absorption cross sections}
%-------------------

As discussed in Sec.~\ref{Sec:II}, the greybody factors for test scalar and vector fields may be determined by solving Eq.~\eqref{eq:Schrodinger} for the form of the field with the boundary conditions~\eqref{eq:transmission}, which are characteristic for the scattering process problem we wish to study. Then, the greybody factor $\Gamma_{\ell}(\Omega)$ of the $\ell$ partial wave with frequency $\Omega$ is expressed as the absorption probability $|A_\ell(\Omega)|^2$. 

%%%%%%%%%%%%%%%%%%%%%%%%%%%%%%%%%%%%%%%%%%%%%%%%%%%%%%%%%%%%%%%%%%%%
\begin{figure*}[t]
\centering
\includegraphics[width=0.49\textwidth]{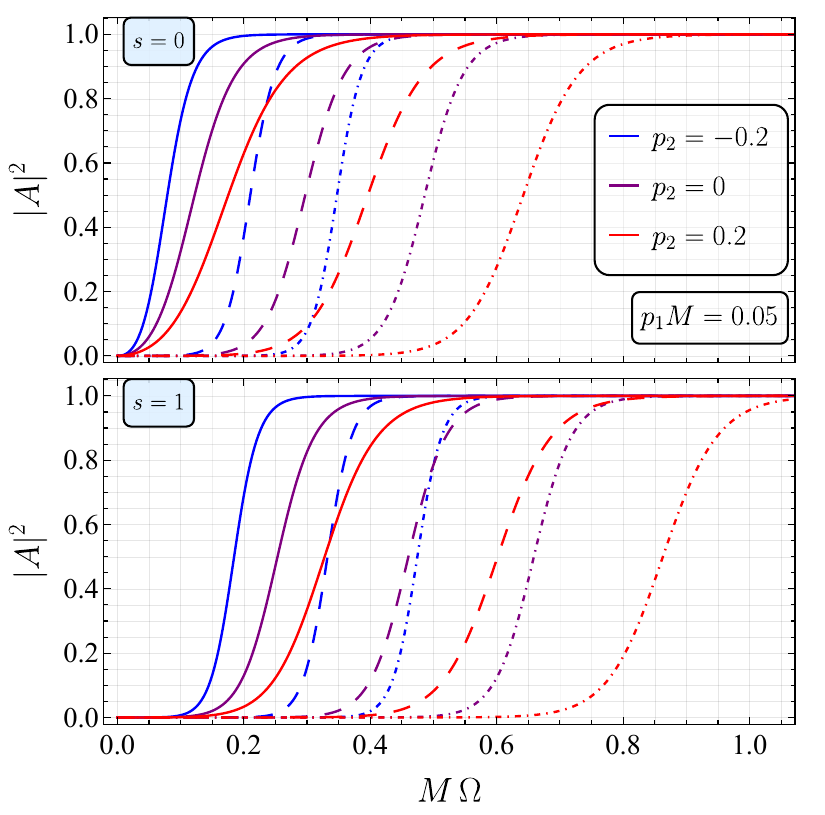}
\includegraphics[width=0.49\textwidth]{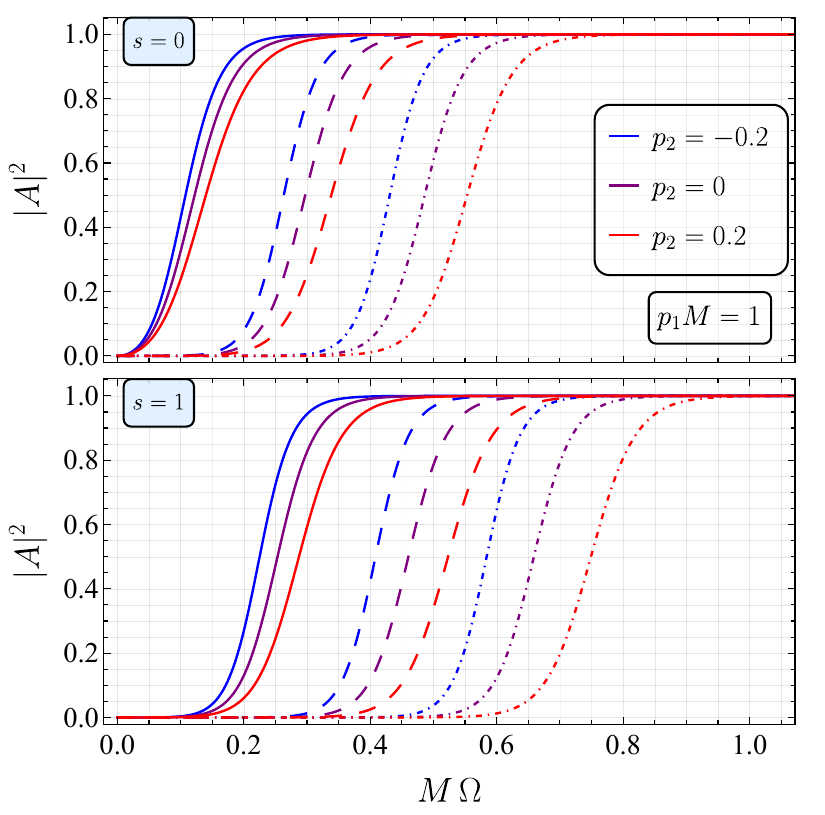}
\caption{Greybody factors for scalar $(s=0)$ and EM $(s=1)$ test fields,~in the background~\eqref{eq:BCKL_metric}, for~$M\,p_1=0.05$ (left panel), and $M\,p_1=1$ (right panel).~Solid curves correspond to $l=s$, dashed to $l=s+1$ and dot-dashed to $l=s+2$.}
\label{fig:GFs_s0_s1}
\end{figure*}
%%%%%%%%%%%%%%%%%%%%%%%%%%%%%%%%%%%%%%%%%%%%%%%%%%%%%%%%%%%%%%%%%%%%

The greybody factors for scalar and vector fields propagating in the background~\eqref{eq:BCKL_metric} of the spherically symmetric black-hole solutions  arising in the context of the parity-symmetric beyond Horndeski theory~\eqref{eq:G_BeyHorn} are presented in Fig.~\ref{fig:GFs_s0_s1}. As it was anticipated due to the form of the effective potentials depicted in Fig.~\ref{fig:Veffs_BCKL}, the greybody factors are suppressed as the parameter $p_2$ increases from negative values to zero and then to positive ones. The suppression is stronger the higher the value $\ell$ of the mode is. As $p_1 M$ takes on larger positive values, and for the case with $p_2 <0$, the greybody factors are further suppressed reaching their maximum values at a larger value of the frequency. Finally, as anticipated, the suppression effect is more dominant for the vectors rather than for scalar fields. 

The relative changes of the greybody factor, being a dimensionless quantity ranging always from zero to unity, are not always easy to discern. The dimensionful quantity of the total absorption cross section, on the other hand is more informative. It is given by (see, e.g.~\cite{Gubser:1997yh, Kanti:2004nr}) 
\beq
\sigma_{\text{abs}}^{(s)}(\Omega)=\sum_{\ell=s}^{\infty}\sigma_\ell^{(s)}(\Omega)\,,
\eeq
where
\beq
\sigma_\ell^{(s)}(\Omega)=\frac{\pi}{\Omega^2}\left(2\ell+1\right)|A_\ell^{(s)}|^2
\label{eq:partial_absorption}
\eeq
are the partial absorption cross sections for each mode of the spin $s$-field, which is easier to study particularly in the low and high-energy regime. We note that the coefficient $(2 \ell +1)$ appearing in Eq.~\eqref{eq:partial_absorption} is due to the multiplicity of states with different angular-momentum orientation number $m$ that correspond to the same partial mode $\ell$ due to the spherical symmetry of the problem.

\begin{figure*}[t]
\centering
\includegraphics[width=0.49\textwidth]{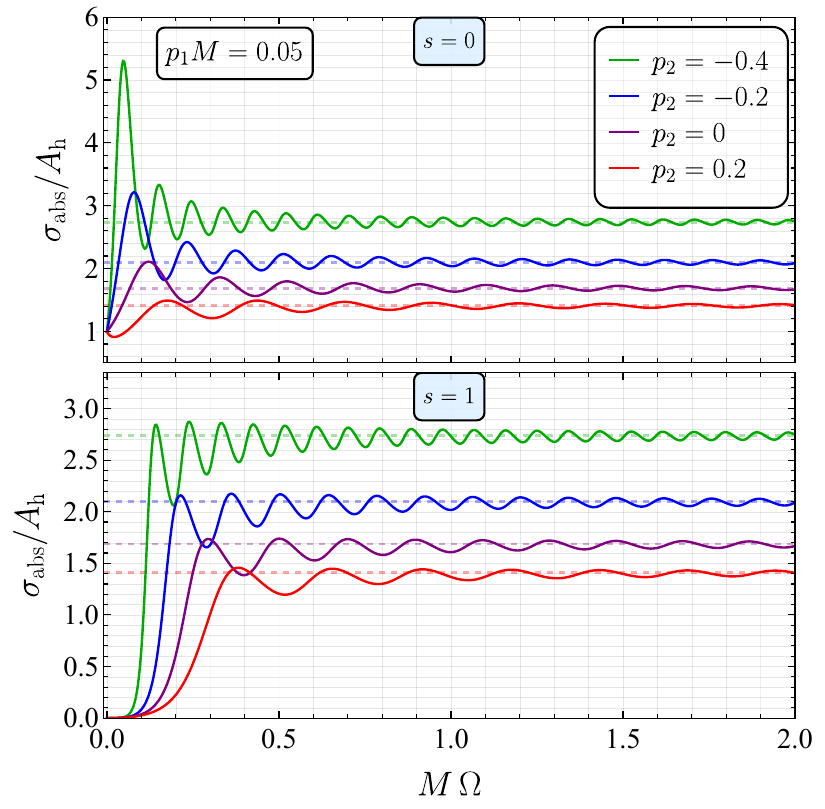}
\includegraphics[width=0.49\textwidth]{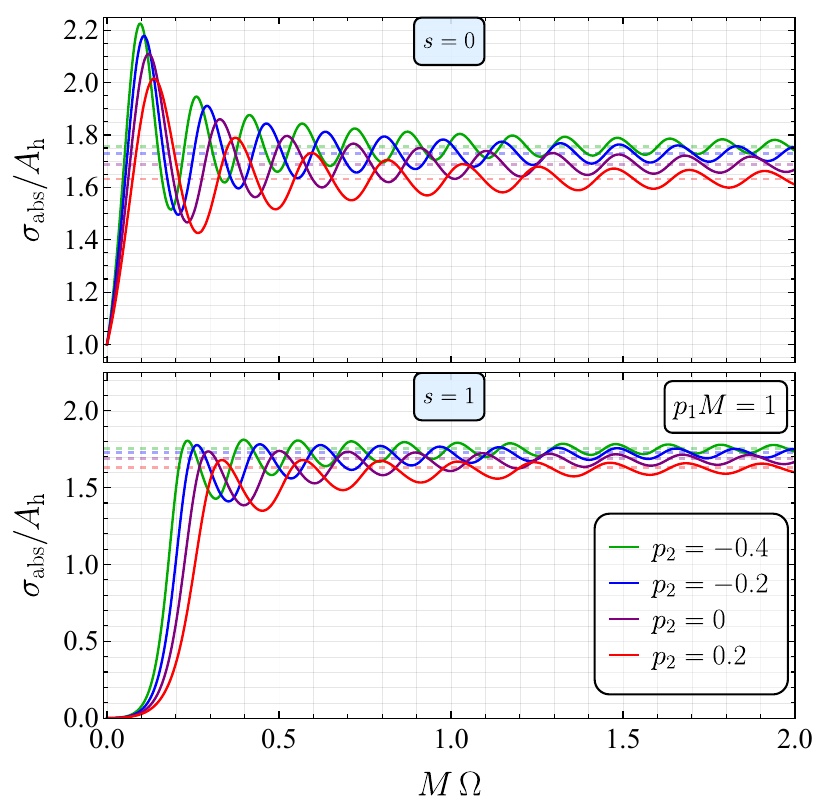}
\caption{Total absorption cross sections normalized by the surface area of the event horizon, for~$M\,p_1=0.05$ (left panel), and $M\,p_1=1$ (right panel), for scalar $(s=0)$ and EM $(s=1)$ test-field perturbations in the background~\eqref{eq:BCKL_metric}. For each total absorption cross section, we have taken into account the partial absorption cross sections with multipole numbers $l \leqslant n+s$, where $n$ is the number of peaks that appear in the figure for each case. The dashed lines correspond to the geometric cross sections as obtained numerically for~\eqref{eq:BCKL_metric}, see also~Table~\ref{tab:sigma_g_Ah}.}
\label{fig:AC_s0_s1}
\end{figure*}

The total absorption cross-section for scalar and vector fields propagating in the background of Eq.~\eqref{eq:BCKL_metric} are presented in Fig.~\ref{fig:AC_s0_s1} normalized to the surface area of the black-hole event horizon $A_h$. We immediately recognize the typical low-energy behavior of the scalar test fields, where the total absorption cross section equals the surface area of the black-hole event horizon (again, see~\cite{Kanti:2004nr} for a general discussion)
\beq
\lim_{\Omega\to0}\sigma_{\text{abs}}^{(0)}(\Omega)=A_h=4\pi r_h^2\,.
\eeq
This behavior holds independently of the values of the parameters $p_1$ and $p_2$. However, this asymptotic value is approached differently as these parameters vary--for example the more negative $p_2$ is, the larger $\sigma_{\text{abs}}^{(0)}$ is in the small-$\Omega$ limit. On the other hand, the total absorption cross-section for vector fields reduces to zero in the low-energy limit as is usually the case. Again, the way this zero value is approached depends mainly on the value of $p_2$ with large negative values enhancing the value of $\sigma_{\text{abs}}^{(1)}$ in the small-$\Omega$ limit.~As a general rule, a rise in the value of $p_1$ suppresses the value of both $\sigma_{\text{abs}}^{(0)}$ and $\sigma_{\text{abs}}^{(1)}$ in the small-$\Omega$ limit. It would be interesting indeed to analytically study the greybody factors and absorption cross-sections to see the explicit dependence of these two quantities on $(p_1, p_2)$.

%%%%%%%%%%%%%%%%%%%%%%%%%%%%%%%%%%%%%%%%%%%%%%%%%%%%%%%%%%%%%%%%%%%%
%\begin{widetext}
%\end{widetext}
%%%%%%%%%%%%%%%%%%%%%%%%%%%%%%%%%%%%%%%%%%%%%%%%%%%%%%%%%%%%%%%%%%%%

Figure~\ref{fig:AC_s0_s1} reveals an interesting behavior also in the high-energy limit: the total absorption cross-sections approach constant asymptotic values which are the same for both scalar and vector fields but vary with the parameters $(p_1, p_2)$. These asymptotic values can be studied in the geometrical optics limit and they are related to the photon sphere radius. A photon propagating in the background~\eqref{eq:BCKL_metric} has an equation of motion of the form
\beq
\left(\frac{1}{r^2}\,\frac{dr}{d\varphi}\right)^2=\frac{1}{b^2} -\frac{A(r)}{r^2}\,,
\label{eq:photon_equation}
\eeq
where $b$ is the ratio of the angular momentum of the particle over its linear momentum. According to the above, the classically accessible regime is given by $b< min(r/\sqrt{A})$, which then defines the photon sphere radius $r_p$, and the geometrical optics value of the total cross-section given by $\sigma_{\rm geo}= \pi b^2_{\rm max}=\pi r_p^2/A(r_p)$.

%------------------------------------------------------
\begin{table}
\centering
\begin{tabular}{| l | c | c | c | c |}
\hline
\multicolumn{5}{|c|}{$p_1\,M=0.05$} \\[1mm]
\hline\hline
$p_2$  & $-0.4$ & $-0.2$ & 0 & 0.2  \\[1mm]
\hline
\text{numerical} & 2.73643 & 2.09676 & 1.6875 & 1.40880 \\
\text{approximation} & 2.73303 & 2.09643 & 1.6875 & 1.40883 \\
\text{ARD} & 0.1242~$\%$ & 0.0155~$\%$ & 0~$\%$ & 0.0021~$\%$ \\
\hline
\hline
\multicolumn{5}{|c|}{$p_1\,M=1$} \\[1mm]
\hline\hline
$p_2$  & $-0.4$ & $-0.2$ & 0 & 0.2  \\[1mm]
\hline
\text{numerical} & 1.75592 & 1.72806 & 1.6875 & 1.63151 \\
\hline
\end{tabular}
\caption{Precise numerical values for the high-frequency limits of the absorption cross section normalized by the total surface area of the event horizon as shown on~Fig.~\ref{fig:AC_s0_s1}.~For~$p_1\,M=0.05$, we also provide their analytical approximations in terms of~\eqref{eq:sig_geo_approx} and the absolute relative difference between the two.}
\label{tab:sigma_g_Ah}
\end{table}
%------------------------------------------------------

%%%%%%%%%%%%%%%%%%%%%%%%%%%%%%%%%%%%%%%%%%%%%%%%%%%%%%%%%%%%%%%%%%%%
\begin{figure*}[t!]
\centering
\includegraphics[width=0.475\textwidth]{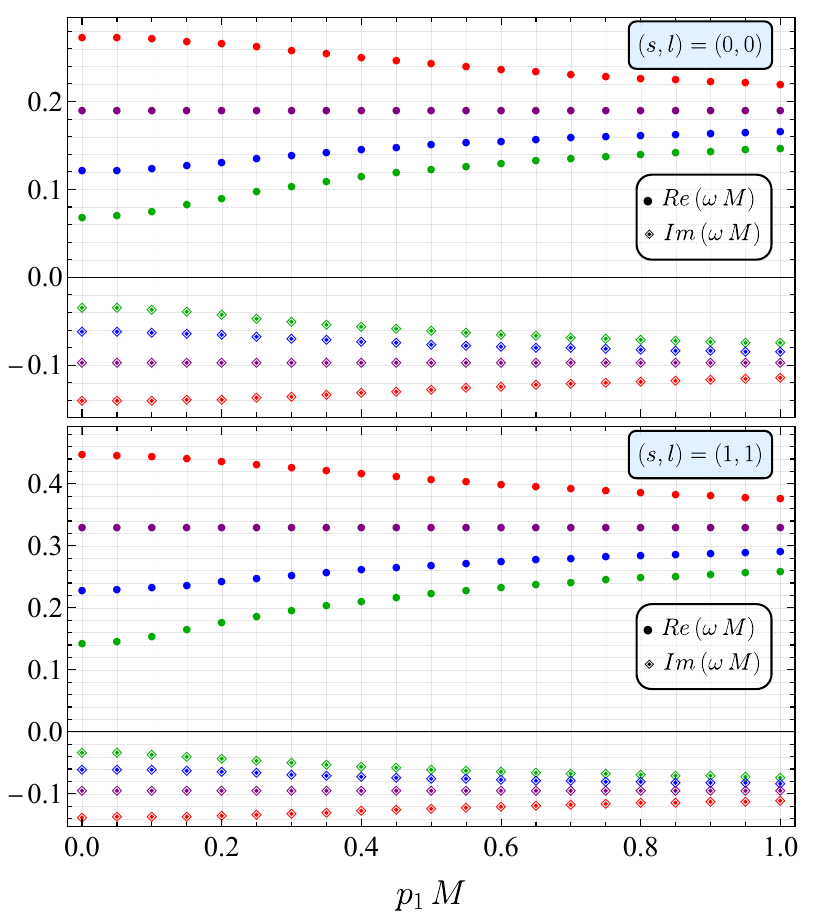}
\includegraphics[width=0.515\textwidth]{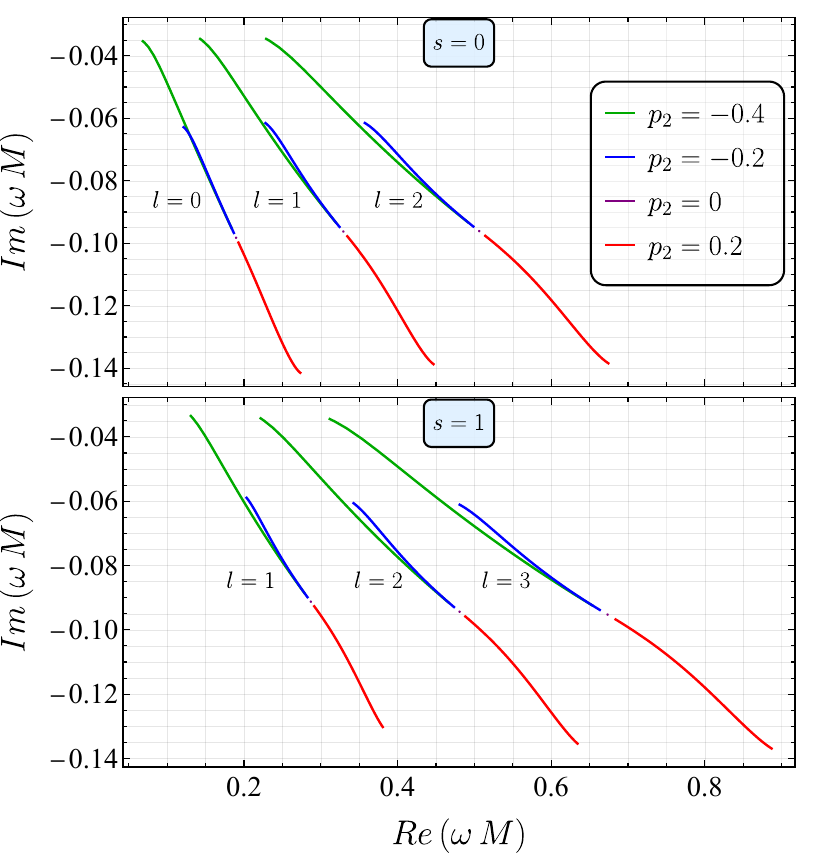}
\caption{Quasinormal modes for scalar $(s=0)$ and EM $(s=1)$ test-field perturbations in the background~\eqref{eq:BCKL_metric}. Left panel, the real and imaginary parts for~$l=s$. Right panel,~$l=s,s+1,s+2$, the curves correspond to a continuous scan of the domain $p_1\,M\in[10^{-4},10]$, with the QNMs approaching the Schwarzschild limit values $(p_2=0)$ monotonically for all $p_2\neq0$ as $p_1\,M\gg1$. The color coding is the same in all panels.}
\label{fig:QNMs_BCKL}
\end{figure*}
%%%%%%%%%%%%%%%%%%%%%%%%%%%%%%%%%%%%%%%%%%%%%%%%%%%%%%%%%%%%%%%%%%%%

The radius $r_p$ corresponding to~\eqref{eq:photon_equation} can be found numerically and its values for some indicative cases of $(p_1, p_2)$ are displayed in Table~\ref{tab:sigma_g_Ah}. As it was clear from Fig.~\ref{fig:AC_s0_s1}, the common high-frequency limit of the total absorption cross section, for both scalar and vector fields, decreases as $p_2$ takes on less negative values and the same holds as $p_1 M$ increases. In an attempt to justify the found behavior, we employ the same approximation as in the derivation of $r_h$. For small values of $p_1 M$, i.e. $p_1\,M  \lessapprox 0.1$, one may derive an analytical approximation for $r_p$. Indeed, using~\eqref{eq:BCKL_p1_approx} we find that the photon sphere radius for~\eqref{eq:BCKL_metric} can be well approximated by
\beq
r_{\rm p} \simeq \frac{3\,M}{1+p_2}\,,
\label{eq:rp_approx}
\eeq
and consequently the geometric cross section is given by
\beq
\sigma_{\rm geo} \simeq \frac{27\pi M^2}{1+p_2\left[3-9M^2 p_1^2+p_2\left( 3+p_2\right)\right]}\,.
\label{eq:sig_geo_approx}
\eeq
Both~\eqref{eq:rp_approx} and~\eqref{eq:sig_geo_approx} reduce to the Schwarzschild values in the GR limit ($p_2 \rightarrow 0$). For nonvanishing values of $p_2$, the approximate values of the total absorption cross-section are given, for the case $p_1 M =0.05$, in the second line of Table~\ref{tab:sigma_g_Ah} together with the absolute relative difference between those values and the exact numerical ones. We observe that, for $p_1 M =0.05$, the agreement is very satisfactory, therefore, the approximation~\eqref{eq:BCKL_p1_approx} indeed suffices for obtaining an accurate analytical approximation for $r_p$. On the other hand, for $p_1 M =1$, the analytical approximation is found, as expected, to be rather inaccurate and thus only the exact numerical values are given.

%%%%%%%%%%%%%%%%%%%%%%%%%%%%%%%%%%%%

%-------------------
\subsubsection{Quasinormal modes}
%-------------------

We now turn our attention to the derivation of the quasinormal modes of the black hole described by Eq.~\eqref{eq:BCKL_metric}.~Solving the Schr\"odinger-type equation~\eqref{eq:Schrodinger} with the appropriate boundary conditions~\eqref{eq:QNM}, we may obtain the complex frequencies $\omega$ for scalar and vector test fields. Following~\cite{Konoplya:2019hlu}, we have used the sixth order WKB method in our analysis.

The real and imaginary parts of the QNM frequencies of the fundamental mode are displayed on the left panel of Fig.~\ref{fig:QNMs_BCKL} in terms of the parameter $p_1\,M$ and for different values of $p_2$. We observe that the real part of $\omega$ gets suppressed whereas the imaginary part becomes less negative as $p_2$ takes on more negative values. This behavior persists for all values of $p_1\,M$~but the deviations from the Schwarzschild values at any given $p_2$ become less pronounced as $p_1\,M$ increases, in agreement with the corresponding limit of~\eqref{eq:BCKL_metric} when $p_1\,r\gg1$. 

On the right panel of Fig.~\ref{fig:QNMs_BCKL}, we depict the QNM curves, $Im (\omega M)$ vs.~$Re (\omega M)$, for the fundamental mode and two higher overtones. We have used the same set of values of $p_2$ and an extended scan of values for $p_1\,M$, i.e. $p_1 M \in [10^{-4}, 10]$. The QNM curves for all three modes extend monotonically from the upper left corner of the plot, for negative values of $p_2$, through the Schwarzschild limit corresponding to the value $p_2=0$, and toward the lower part of the plot, where both $|Im (\omega M)|$ and $Re (\omega M)$ increase further as $p_2$ takes on positive values. The same behavior is observed for both scalar and vector fields.

%===============================================
\section{Analysis with backreaction}
\label{Sec:IV}
%===============================================

In this Section, we will go beyond the test-field analysis and will look into the full theoretical framework under linear perturbations for two classes of Horndeski solutions (see Secs.~\ref{Sec:IV} A and B). To our knowledge, this is the first time that someone studies GB factors of non-GR black holes by taking into account the backreaction of the additional field(s) into the metric elements.

If we consider standard GR supplemented by one real scalar field, the metric and scalar perturbations at the linear level are given by
\begin{align}
g_{\mu\nu}&=g_{\mu\nu}^{(0)}+\varepsilon\, \delta g_{\mu\nu}\,,\\
\phi&=\phi^{(0)}+\varepsilon\,\delta\phi\,,
\end{align}
where for spherically symmetric configurations the perturbations can be expanded as
\begin{align}
    \delta g_{\mu\nu}(t,r,\theta,\varphi)=& \int d\omega\, h_{\mu\nu}(r) Y_{\ell}^{m}(\theta,\varphi)\,e^{-i\omega t}\,,\label{eq:metric_decomposition}
    \\[2mm] 
    \delta\phi(t,r,\theta,\varphi)=& \int d\omega\, \frac{\phi_1(r)}{r}Y_{\ell}^{m}(\theta,\varphi)\,e^{-i\omega t}\, .\label{eq:scalar_decomposition}
\end{align}
Studying QNMs in such systems reveals significant differences between the axial and polar sectors.
To begin with, there is a spectral asymmetry between the two parities which is absent in GR.
Moreover, two types of modes exist in the polar sector, i.e., \textit{grav-led} modes and \textit{scalar-led} modes, which are distinguished based on the type of perturbation that gets excited (gravitational or scalar).

In this work, we will limit our computations to the axial sector, as in this case only the gravitational-led modes are present and the scalar perturbation decouples.
In the Regge-Wheeler gauge, we have the following decomposition of the metric perturbations in the axial sector:
\begin{equation}
    h_{\mu\nu}^{\text{axial}}=
    \begin{bmatrix}
    0 & 0 & -h_0\csc\theta\,\partial_\varphi & h_0\sin\theta\,\partial_\theta\\
    0 & 0 & -h_1\csc\theta\,\partial_\varphi & h_1\sin\theta\,\partial_\theta\\
    \text{Sym} & \text{Sym} & 0 & 0\\
    \text{Sym} & \text{Sym} & 0 & 0
    \end{bmatrix}\, .
\end{equation}
This allows us to derive a single master equation for the axial perturbations, which may be brought to a Schr\"odinger-type form
\begin{equation}
    \frac{d^2Q}{dr_*^2}+\left[\omega^2-V\right]Q=0\,.
\label{eq:master_eq}
\end{equation}
In the above, the tortoise coordinate is defined via the relationship
\begin{equation}
    \frac{dr_*}{dr}\equiv \sqrt{g(r)}\,,
\label{eq:tortoise}
\end{equation}
where the function $g(r)$ depends on the exact model we study, and so is the function $f(r)$ that we use to define the master perturbation function
\begin{equation}
    Q(r)=f(r)\,h_1(r)\,.
\end{equation}
Notice that we choose to remove the perturbation function $h_0$ by making use of the $(\theta \varphi)$ component of the Einstein equations. This is simply a matter of choice and we could have equivalently chosen to rid our analysis of $h_1$ leading to a different definition for $Q(r)$.

In order to calculate the QNMs and GB factors in the case of theories with backreaction, we solve the aforementioned Eq.~\eqref{eq:master_eq} with the appropriate boundary conditions.
For the QNMs, in particular, it is usually simpler to solve the system of two coupled first order differential equations (with $h_0$ and $h_1$ being the perturbation functions) rather than working with the master equation. In any case, we have confirmed that the results are consistent between the two approaches.

We now present the general methodology used to solve for QNMs and greybody factors in such systems of coupled differential equations. Let us assume a system of $N$ coupled differential equations--in our case $N=1$ or $N=2$ depending on whether we work with the master function or $(h_0,h_1)$--and define the perturbation vector $\boldsymbol{\Psi}=\big(\Psi^{(1)}\ldots\Psi^{(N)}\big)$.
Assuming that the perturbation functions near the event horizon are regular, we consider the following near-horizon expansion:
\begin{equation}
\Psi^{(j)}=\;(r-r_h)^b\sum_{n=0} \psi_n^{(j)}(r-r_h)^{n+n_{j}}\,,\label{eq:bchor}
\end{equation}
where $j=1,\ldots N$ and $n_{j}$ depends on which perturbation function $\Psi^{(j)}$ we are considering, while $b=-i\omega\sqrt{a^{(1)}b^{(1)}}$, where $a^{(1)},b^{(1)}$ are the first order expansion coefficients at the horizon for the background solution, as those are defined from Eqs. \eqref{eq:expA}-\eqref{eq:expB}.
For the study of the axial sector in particular for $\boldsymbol{\Psi}=\big(h_0,h_1\big)$ we have $n_{1}=0$ and $n_{2}=-1$.
Similarly to the treatment of the background expressions, substituting the above in the field equations allows us to determine the coefficients in an order by order approach.
At infinity we have in principle two contributions, from an outgoing and an ingoing solution.
\begin{equation}
    \Psi^{(j)}\sim A^{(j)} e^{-i \omega r}r^{q+m_{j}}+B^{(j)} e^{i \omega r}r^{q+m_{j}}\,. \label{eq:bcfar}
\end{equation}
The two aforementioned expressions for $\Psi^{(j)}$, Eqs.~\eqref{eq:bchor}-\eqref{eq:bcfar}, serve as our boundary conditions for the system of coupled equations.
The constants $q$ and $m_j$ depend on the problem at hand similarly to $n_j$.
For a modified background $q=-i\omega(\tilde{a}^{(1)}+\tilde{b}^{(1)})/2$ where $\tilde{a}^{(1)},\tilde{b}^{(1)}$ are the first order expansion coefficients at infinity for the background solution defined in Eqs.~\eqref{eq:expAi}-\eqref{eq:expBi}.
For the axial perturbations we have $m_{1}=m_{2}=1$.
In the implementation of our shooting method for QNMs, where we choose to work with both $h_0$ and $h_1$, we construct the appropriate $2\times 2$ matrix whose elements correspond to the coefficients $B^{(j)}$, and then demand that its determinant vanishes when $\omega$ becomes a quasinormal frequency.

%------------------------------------%
%------------------------------------%
\subsection{Shift-symmetric Horndeski theory}
%------------------------------------%
%------------------------------------%
Shift symmetric Gauss-Bonnet gravity has been used extensively as an example framework in performing various phenomenological studies and numerical relativity simulations primarily by focusing on the so-called minimal model, i.e., the one containing only the hair-sourcing scalar-GB interaction.

However, when the shift-symmetric Horndeski theory beyond the minimal model was studied, it was shown that a plethora of interesting characteristics associated with the additional interactions, which are absent in the minimal scenario, may be present \cite{Thaalba:2022bnt}.
Moreover, a full analysis of the axial and polar QNM spectrum within this framework was performed in~\cite{Antoniou:2024hlf}, and interestingly nontrivial deviations from GR were highlighted.
Here, we will expand on that work by considering the GB factors, focusing on the easier-to-handle axial sector, as explained above.

The action functional of the theory is given by
\begin{equation}
\begin{split}
    S=\frac{1}{2k}\int \mathrm{d}^4& x\,\sqrt{-g}\bigg[\frac{R}{2}+X+\alpha\,\phi\,\mathcal{G}\\
    &+\gamma \,G_{\mu\nu}\nabla^\mu\phi\nabla^\nu\phi
    +\sigma X\Box\phi +\kappa\, X^2 \bigg],
\end{split}
\label{eq:action}
\end{equation}
and may be obtained from action~\eqref{eq:framework} by selecting
\begin{equation}
\begin{split}
    &
    G_2(X) \coloneqq X+\kappa X^2\quad , \quad
    G_3(X) \coloneqq -\sigma X\, ,\\
    &
    G_4(X) \coloneqq 1/2 + \gamma X\quad , \quad
    G_5(X) \coloneqq -4\alpha\ln|X|\,,
\end{split}
\label{eq:Gi_functions}
\end{equation}
setting the beyond-Horndeski contributions to zero. In the above, $(\alpha, \gamma, \sigma, \kappa)$ are coupling parameters of the theory, and $\mathcal{G}$ stands for the Gauss-Bonnet quadratic gravitational term.

\begin{figure}[t]
    \centering
    \includegraphics[width=1\linewidth]{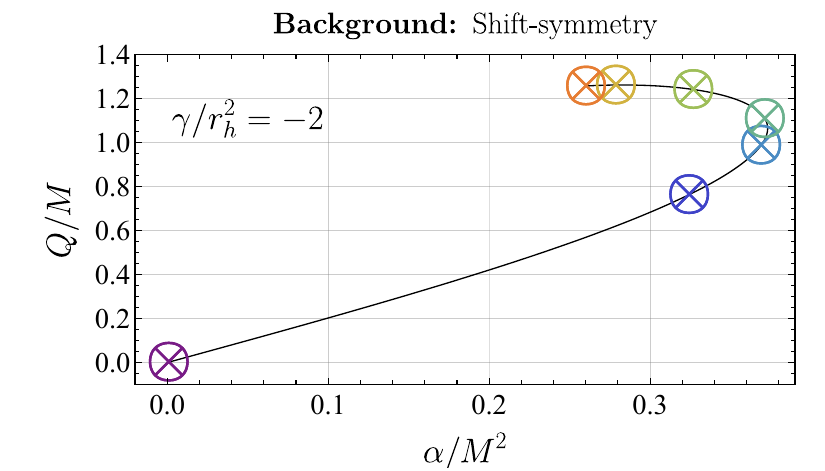}
    \caption{Scalar charge of the hairy solutions in the shift-symmetric theory, versus the hair-sourcing parameter $\alpha$. Each point in the solid line corresponds to a different black hole solution. We also show the seven points we use as reference in our analysis.}
    \label{fig:BG_shift_symmetry}
\end{figure}

%%%%%%%%%%%%%%%%%%%%%%%%%%%%%%%%%%%%%%%%%%%%%%%%%%%%%%%%%%%%%%%%%%%%
\begin{figure*}[t!]
\centering
\includegraphics[width=0.49\textwidth]{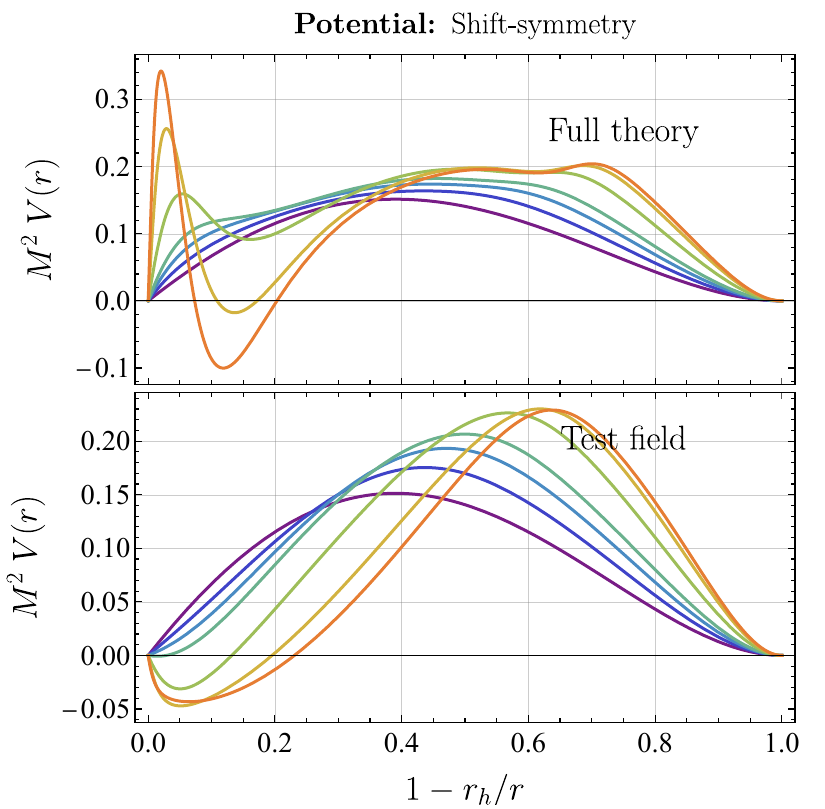}
\includegraphics[width=0.49\textwidth]{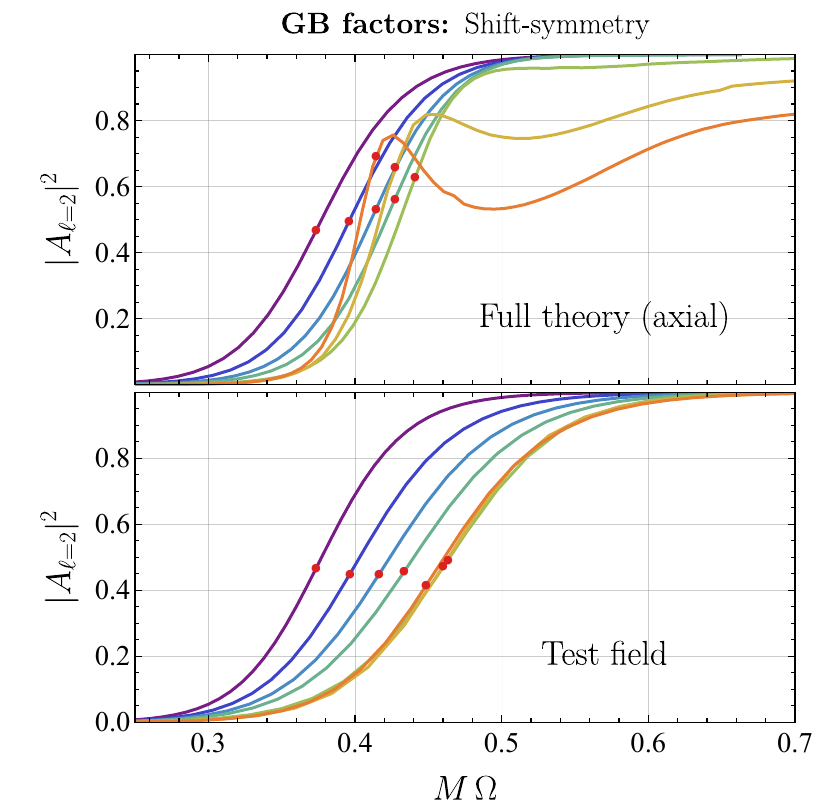}
\caption{{\it Left panel:} The perturbations potential in the shift-symmetric theory, for the seven chosen reference points (black-hole solutions) in the case of the full theory (axial perturbations), and of a test field. {\it Right panel:} Greybody factors for the $\ell=2$ scenario in the shift-symmetric theory, both in the full theory and for the test field. The red dots denote the real part of the corresponding QNM frequency.}
\label{fig:V_shift_symmetry}
\end{figure*}
%%%%%%%%%%%%%%%%%%%%%%%%%%%%%%%%%%%%%%%%%%%%%%%%%%%%%%%%%%%%%%%%%%%%

In what follows, we choose to work with the value $\gamma/r_h^2=-2$. For a value of this order it has been shown that the existence lines for the hairy black holes as well as the QNM spectrum demonstrate a nonmonotonic behavior \cite{Thaalba:2022bnt,Antoniou:2024hlf}, which will hopefully provide a deeper insight in the GB factors and their relationship with the QNMs. In Fig.~\ref{fig:BG_shift_symmetry}, we show the aforementioned existence line, i.e., the dependence of the black-hole scalar charge on the hair-sourcing parameter $\alpha$.
After choosing to work with $\gamma/r_h^2=-2$, we pick seven different points along the hairy black-hole existence line. Each one of the seven points corresponds to a different choice of the hair-sourcing parameter $\alpha$. The GR limit is retrieved for $\alpha\to 0$ and corresponds to the reference point on the lower left part of the existence line. The final point at the upper part of the curve is chosen so that $\alpha/r_h^2$ acquires its maximum allowed value--any further increase would result in a finite-radius singularity larger than the black hole radius or a violation of the hyperbolicity of the master equation, whichever comes first [see~\eqref{eq:condition_1} and  \eqref{eq:condition_2}].

We then proceed to determine the potential that appears in the Schr\"odinger-type equation~\eqref{eq:master_eq} for the perturbations in the axial sector. Albeit cumbersome, this task can be performed and the axial potential $V$ is identified. In the left panel of Fig.~\ref{fig:V_shift_symmetry}, in the upper plot, we show the potential for each of these seven points (seven black-hole solutions) in the axial sector of the full theory. For comparison, in the lower plot, we also present the potential for a spin-2 field that is not affected by the scalar-field perturbations. In this case, the form of the potential follows from Eq.~\eqref{eq:Schrodinger} for $s=2$. The differences in the form of the potential for the two types of perturbations, for the same black-hole solutions, are profound over the whole radial regime. We also observe that, for both types, the single-peak form of the potential for solutions corresponding to small values of $\alpha/M^2$ is replaced by a form with multiple extrema, for solutions arising for large values of $\alpha/M^2$; this form exhibits potential wells and, in the case of the full theory, multiple peaks. We note that the tortoise coordinate employed in the Schr\"odinger-type equation~\eqref{eq:master_eq} is defined through~\eqref{eq:tortoise} where
\begin{equation}
    g(r)=\frac{1-4 \alpha  B' \phi'-B (8 \alpha  \phi''+\gamma  \phi'^2)}{B (A-4 \alpha  B A' \phi'+\gamma  A B \phi'^2)}\, .
\label{eq:condition_1}
\end{equation}
Demanding that the quantity above is nonzero in our regime of computations yields an upper limit for $\alpha$. In particular, for the choice $\gamma/r_h^2=-2$, we find $\alpha_\text{max}/r_h^2\approx 0.34$. 

In the right panel of Fig.~\ref{fig:V_shift_symmetry}, we present the corresponding GB factors for the shift-symmetric theory~\eqref{eq:action}. The upper plot depicts the GB factors for the gravitational perturbations in the axial sector of the theory, whereas the lower plot shows the GB factor for a test spin-2 field, to allow for a comparison. These follow by solving Eqs.~\eqref{eq:master_eq} and~\eqref{eq:Schrodinger}, respectively, under the appropriate boundary conditions. 
We observe that, for both types of perturbations and for solutions with small values of $\alpha/M^2$, the GB factors retain their monotonic behavior and get suppressed as $\alpha/M^2$ increases. In the case of the test field, the presence of the well in the form of the potential, for large values of $\alpha/M^2$, results in a constraining of the anticipated suppression of the GB factors. This could potentially be interpreted in light of the findings of a recent study on the general response of greybody factors to theory-agnostic deformations of the effective potential~\cite{Konoplya:2025ixm}, where it was demonstrated that near-horizon wells in the potential tend to enhance the greybody factors at least in the low-frequency regime. The observed constraining of the suppression of the greybody factors could be a consequence of the balancing between the enhancement due to the deepening of the near-horizon well and the simultaneous increase of the height of the effective-potential barrier as $\alpha/M^2$ increases. For the gravitational perturbations, on the other hand, the multiple-peak form of the potential is directly reflected in a similar form of the greybody factor curves. In this case, the potential well leads to an enhancement of the GB factors at the intermediate frequency regime whereas the multiple barriers cause the GB factors to reach their maximum value at a much higher frequency of the propagating mode. These observations seem to once again be in qualitative agreement with the simple theory-agnostic test-field analysis of~\cite{Konoplya:2025ixm} where it was found that near-horizon bumps (positive height deformations) on the effective potential tend to suppress the greybody factors, particularly in the low-frequency regime, while dips (negative height deformations) enhance them. Then, the combined effect of the two could result in a frequency profile for the greybody factors similar to the one observed in Fig.~\ref{fig:V_shift_symmetry}.

\begin{figure}[t]
    \centering
    \includegraphics[width=1\linewidth]{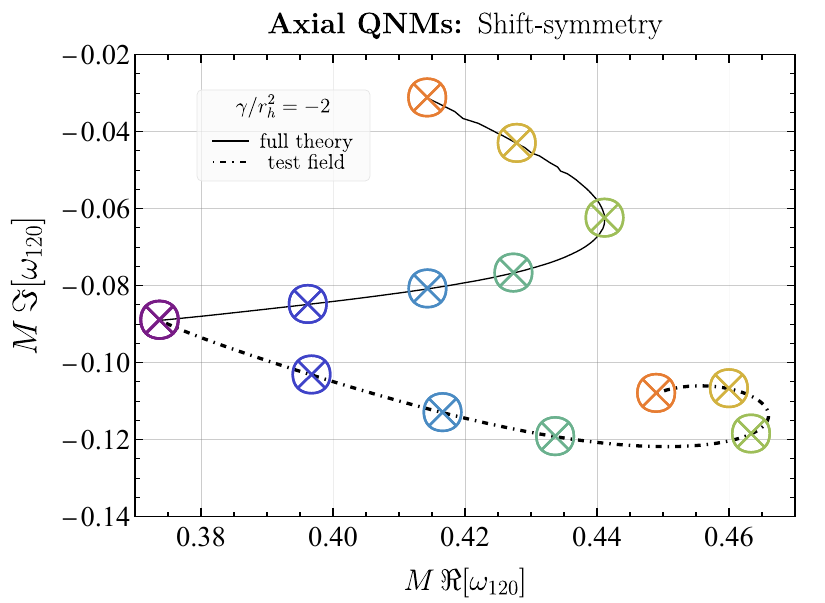}
    \caption{$\ell=2$ QNM frequencies derived in shift symmetric Horndeski gravity with the choice $\gamma/r_h^2=-2$. The solid line corresponds to the axial sector of the full theory, while the dashed one to the modes describing a test field.}
    \label{fig:QNMs_shift_symmetry}
\end{figure}

Let us finally discuss the quasinormal modes in this theory.  Solving again Eqs.~\eqref{eq:master_eq} and~\eqref{eq:Schrodinger} under the purely incoming-purely outgoing boundary conditions, we derive the quasinormal frequencies for the axial gravitational and spin-2 test field perturbations, respectively. In Fig.~\ref{fig:QNMs_shift_symmetry}, we show the real and imaginary parts of the QNMs for these two types of perturbations. 
Both QNM curves emanate from the GR limit on the leftmost side of the plot, but follow distinctly different paths.
The solid line for the axial gravitational perturbations is consistent with the results presented in~\cite{Antoniou:2024hlf}. In this case, an increase in the value of the hair-sourcing parameter $\alpha/M^2$ leads to a decrease in the negative imaginary part of the QNM frequency and to an overall increase of its real part; the first characteristic points to a reduced stability of the corresponding solutions under gravitational perturbations, compared to that of their GR analog. The test-field QNM curve, on the other hand, exhibits an increase in both the imaginary and the real part of the QNM frequency for all hairy black-hole solutions, compared to the Schwarzschild solution. For both types of fields, we also observe that the QNM curves present a nonmonotonic form on their far-right part where the maximum value of $\alpha/M^2$ is gradually approached; the reference points that lie in this regime were shown above to be associated with a potential profile with multiple extrema that led to a modified behavior of the greybody factors. Here, we see that the QNM frequencies, both the real and imaginary parts, are also affected by this feature. In order for this connection to be more transparent, the real parts of the quasinormal frequencies are denoted with red dots in the corresponding plots of the greybody factors on the right panel of Fig.~\ref{fig:V_shift_symmetry}. From this, it is clear that any deviation from the traditional single-peak form of the potential, affects simultaneously the greybody-factor curves and the quasinormal frequencies.

%------------------------------------%
%------------------------------------%
\subsection{Quadratic-quartic-scalar-GB theory}
%------------------------------------%
%------------------------------------%

We now turn to an additional subtheory within the Horndeski framework, which allows for scalarized black hole solutions, described by the following action functional
\begin{equation}
\label{eq:Action}
\begin{split}
    S=\frac{1}{2k}\int \mathrm{d}^4x & \,\sqrt{-g}\bigg[ R + X + \left(\frac{\alpha}{2} \phi^2+\frac{\zeta}{4}\phi^4 \right)\mathcal{G}\bigg].
\end{split}
\end{equation}
The above action accommodates a combination of quadratic and quartic coupling functions of the scalar field to the Gauss-Bonnet term $\mathcal{G}$. As was demonstrated in~\cite{Antoniou:2024gdf}, the quadratic coupling is actually the one that supports the scalar hair of the black-hole solutions that arise in the context of this theory. However, these solutions would be unstable were it not for the existence of the quartic coupling in the theory: stable spontaneously scalarized black holes exist only when $\zeta/\alpha\lesssim -0.7$.
Unlike the shift-symmetric theory of the previous subsection, here hairy solutions bifurcate from the GR branch for a nonzero value of the hair-sourcing parameter $\alpha$, namely $\alpha/M^2\approx 0.73$.

%%%%%%%%%%%%%%%%%%%%%%%%%%%%%%%%%%%%%%%%%%%
\begin{figure}[t!]
    \centering
    \includegraphics[width=1\linewidth]{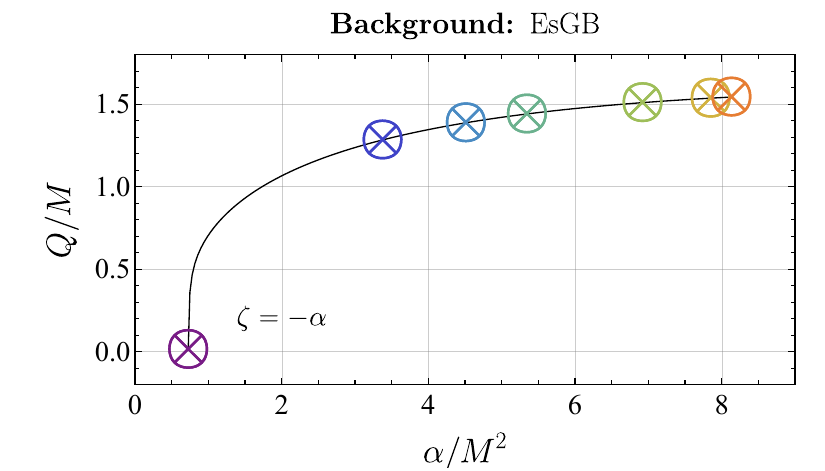}
    \caption{Scalar charge of the hairy solutions in the quartic sGB theory, versus the hair-sourcing parameter $\alpha$. Each point in the solid line corresponds to a different black-hole solution. We also show the seven reference points we use in our analysis.}
    \label{fig:domain_EsGB}
\end{figure}
%%%%%%%%%%%%%%%%%%%%%%%%%%%%%%%%%%%%%%%%%

The function $g(r)$ determining the tortoise coordinate through~\eqref{eq:tortoise} is now given by
\begin{equation}
    g(r)=\frac{1-2 \phi  \phi ' B'-4 \phi  B \phi '' -4 B (\phi ')^2}{B(A (\alpha +\zeta  \phi ^2)^{-1}-2 \phi  B \phi ' A')}\,,
\label{eq:condition_2}
\end{equation}
and sets the upper limit for the hair-sourcing parameter at $\alpha/M^2\approx 8.13$.

The spectrum of quasinormal modes for these scalarized black-hole solutions was derived in~\cite{Antoniou:2024gdf} for indicative values of the parameter $\zeta/\alpha$ in the stable regime, and the modifications from the GR limit were studied in general. Here, we will choose a particular value of this parameter, i.e. $\zeta/\alpha=-1$, which lies in the stable part of the parameter space, and perform a comprehensive study of the family of black-hole solutions that emerge, along the lines of the previous subsection.

We begin with the existence line of the family of the corresponding black-hole solutions, which is depicted in Fig.~\ref{fig:domain_EsGB}. The scalar charge of the solutions is given in terms of the hair-sourcing parameter $\alpha/M^2$. The existence line in this case extends from the far left point, which corresponds to a fixed minimum value of $\alpha/M^2$, to the far right, which corresponds to the maximum allowed value of $\alpha/M^2$. Again, seven points have been chosen along this line as reference points.

%%%%%%%%%%%%%%%%%%%%%%%%%%%%
\begin{figure}[t]
\centering
\includegraphics[width=0.49\textwidth]{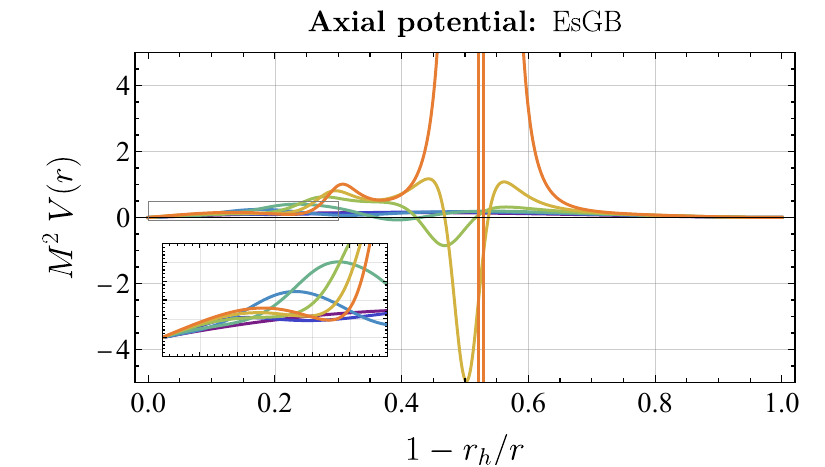}
\includegraphics[width=0.49\textwidth]{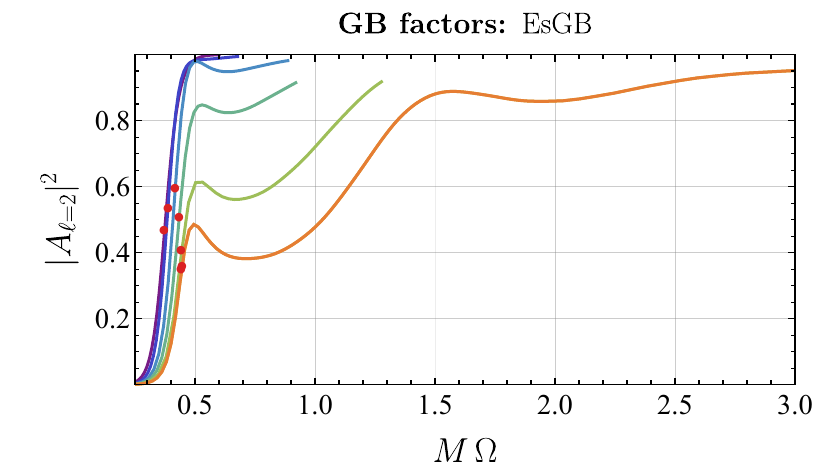}
\caption{\textit{Top:} the potential for gravitational perturbations in the EsGB theory, in the case of the full theory (axial sector) for the seven chosen reference points.
\textit{Bottom:}  greybody factors for the $\ell=2$ modes in the axial sector of the full EsGB theory. The red dots denote the real part of the corresponding QNM frequency.}
    \label{fig:BG_V_EsGB}
\end{figure}
%%%%%%%%%%%%%%%%%%%%%%%%%%%%%%%%%%%%

In light of the results presented in the previous subsection, here we discard the study of the spin-2 test field as nonaccurate, and focus only on the full-theory analysis and on the axial sector of perturbations. The form of the corresponding potential for the seven chosen black-hole solutions is depicted in the top panel of Fig.~\ref{fig:BG_V_EsGB}. A form with multiple extremal points is again observed, similar to the one found in~\cite{Antoniou:2024gdf}, which, in this theory, is actually prominent in almost every curve apart from the one corresponding to the lowest value of the hair-sourcing parameter $\alpha/M^2$. As this parameter increases, a well and multiple peaks appear; these features exhibit a tremendous increase in their depth or height as we move toward the upper part of the existence line. 

The multiple-extrema form of the potential is expected to provide another interesting case for the study of the corresponding GB factors.~These quantities are presented in the lower panel of Fig.~\ref{fig:BG_V_EsGB}. The deviation from the typical smooth curves extending monotonically between zero and unity, an example of which was presented in the test-field approximation of Sec.~\ref{Sec:III}, is more than evident. The greybody curves, for the seven chosen solutions, do not show significant suppression in the low-energy regime as the hair-sourcing parameter $\alpha/M^2$ increases, however, they exhibit a radically different behavior in the high-energy regime; there, the combination of well and peaks in the potential lead to the emergence of multiple extrema and to an overall suppression of the GB curves, especially so as we reach the largest values of $\alpha/M^2$.

\begin{figure}[t]
    \centering
    \includegraphics[width=1\linewidth]{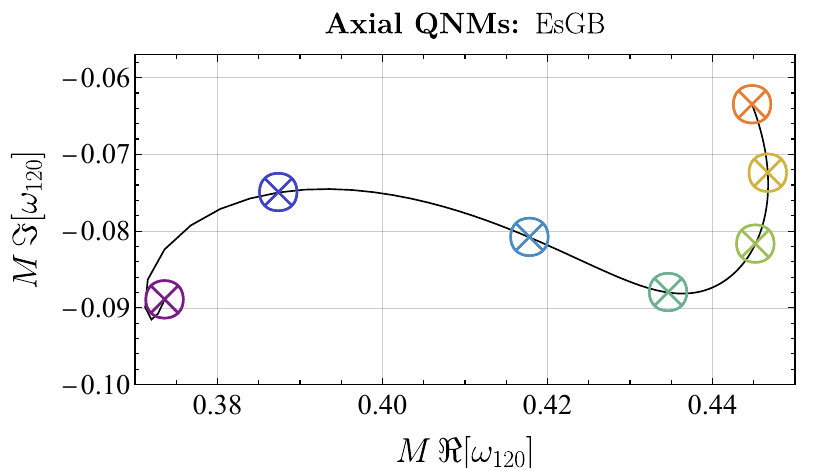}
    \caption{$\ell=2$ QNM frequencies derived in the EsGB model with $\zeta/\alpha=-1$. The seven reference black hole solutions are denoted with different colors.}
    \label{fig:QNMs_GBf_EsGB}
\end{figure}

Finally, in Fig.~\ref{fig:QNMs_GBf_EsGB}, we present the quasinormal curve for the complete existence line of the black-hole solutions emerging for the indicative choice of $\zeta/\alpha=-1$, with the seven reference points shown again. The effect of the multiple extrema of the potential is also evident here: as $\alpha/M^2$ increases from its lowest to its maximum value, both the imaginary and real parts of the quasinormal frequency of the fundamental mode show areas of both decrease and increase. Comparing the results for the two solutions corresponding to the smallest and largest value of $\alpha/M^2$, we conclude that black holes with the highest hair-sourcing parameter are characterized by a larger real part and a smaller (less negative) imaginary part of their quasinormal frequency.

%%%%%%%%%%%%%%%%%%%%%%%%%%%%%%%%%%%%%
%%%%%%%%%%%%%%%%%%%%%%%%%%%%%%%%%%%%%%%%%

%===============================================
\section{Conclusions}
\label{Sec:Conclusions}
%===============================================

The beyond Horndeski action contains a wealth of scalar-tensor theories parametrized by the $G_i$ and $F_i$ coupling functions, which provide the framework for the potential discovery of new black-hole solutions with modified characteristics compared to traditional GR. These characteristics may include the greybody factors, associated with the scattering process that a field undergoes in the vicinity of a black hole, and the quasinormal frequencies, that characterize the way a field or the spacetime itself responds to a perturbation. 

Both types of these physical quantities may be derived by following two different approaches.~In the first approach, the test field (a scalar or a vector one, in the context of our analysis) is assumed not to affect the scalarized black-hole solution, and thus to propagate in a predetermined, fixed gravitational background. The second approach is followed when we turn our attention to the gravitational field itself, and consider perturbations not only to the metric but also to the scalar degree of freedom it couples to, in the context of the Horndeski theory.  

In the present analysis, we have adopted both approaches and applied them in turn, depending on the setting, to a variety of static, spherically symmetric, scalarized black-hole solutions arising in the context of the general framework of beyond Horndeski theory.  

To this end, we first considered the analytic form of a scalarized black hole solution~\cite{Bakopoulos:2022csr} emerging in the parity-symmetric sector of the beyond Horndeski theory. This solution is parametrized by two coupling parameters $p_1$ and $p_2$ and assumes, under the condition that $p_2<0$, the form of a Reissner-Nordstr\"om black hole at asymptotic infinity. In this fixed background, we considered the propagation of massless test scalar and electromagnetic fields, and solved the corresponding equations of motion. 

Addressing first the case of a scattering process, we computed the greybody factors for the two types of test fields. The observed behavior of these quantities was in sheer accordance with the form of the effective potentials, which in this case have the form of single-peak gravitational barriers, and provided useful information regarding the role of the two coupling parameters of the theory. In particular, we found that as the parameter $p_2$ takes on more negative values, the height of the barrier decreases, and the greybody factors are accordingly enhanced. In that regime, the enhancement effect becomes more profound as the positive parameter $p_1 M$ decreases, too. The same pattern is observed for the total absorption cross section for both test fields and over the whole frequency regime. Given that the specific scalarized solution deviates more from the Schwarzschild solution as $p_2$ becomes more negative and $p_1 M$ gets smaller, we conclude that this scalarized black hole could be characterized by distinctly larger values of the absorption probabilities compared to those for the Schwarzschild solution. It would be indeed interesting to compute the exact spectra of Hawking radiation from this--and similar--scalarized black hole(s). 

The study of the quasinormal frequencies for scalar and vector test fields in the same black-hole background, both for the fundamental and higher modes, revealed in turn that the real part of $\omega$ gets suppressed, whereas the imaginary part becomes less negative as $p_2$ takes on more negative values. As a result, non-GR solutions of this type tend to have smaller characteristic oscillation frequencies and larger damping times the more they deviate from the Schwarzschild form; this could also provide an additional distinct feature of non-GR solutions.

The approach of the full-theory analysis was adopted next as we turned our attention to the perturbations of the gravitational field itself. We considered linear perturbations in the form of both the metric and scalar field, and derived a decoupled master equation for the axial sector of the gravitational perturbations, which nevertheless takes into account the backreaction of the scalar field into the metric elements.

Employing the numerically determined scalarized black hole solutions of the shift-symmetric Horndeski theory as the background solution, we first computed the greybody factors associated with the axial sector. We chose to work with the value $\gamma/r_h^2=-2$ as, in this area of the parameter space of the theory, the existence line, i.e.~the dependence of the black-hole scalar charge on the hair-sourcing parameter $\alpha$, presents a nonmonotonic behavior. In this existence line, we also chose seven different black-hole solutions, each one corresponding to a different choice of the hair-sourcing parameter $\alpha$ with the GR solution arising for $\alpha=0$. 

Our subsequent analysis indeed confirmed the deviation of the form of the gravitational potential from the traditional form of a single-peak barrier, especially as the value of the hair-sourcing parameter $\alpha/M^2$ increased. 
Indeed, as $\alpha/M^2$ approached its maximum value, the effective potential exhibited multiple peaks and wells. These features are also transferred to the form of the GB factors with the wells leading to an enhancement and the multiple barriers to a suppression and, therefore, a delay in reaching their maximum value.   

The study of the quasinormal frequencies for the same black-hole solutions of the shift-symmetric theory showed that an increase in the value of the hair-sourcing parameter $\alpha/M^2$ leads to an overall increase of the real part of the QNM frequency and to a decrease in the (negative) imaginary part, as was also shown in \cite{Antoniou:2024hlf}. Therefore, also here the deviation from GR leads to distinctly different oscillation frequencies and larger damping times for these oscillations. We should also note here that the test-field analysis, performed for the sake of comparison, revealed significant differences in the values of QNMs from the full-theory analysis, thus justifying the need for the latter in any realistic study. However, both analyses were able to confirm that any deviation from the traditional single-peak form of the potential affects simultaneously the greybody-factor curves and the quasinormal frequencies.

We finally turned to a subclass of the Horndeski theory which accommodates a combination of quadratic and quartic coupling functions of the scalar field to the Gauss-Bonnet term, with the latter stabilizing the scalarized black-hole solutions supported by the former. We worked with the choice $\zeta/\alpha=-1$ which lies in the stable part of the parameter space, and studied in detail the whole existence line of the black-hole solutions of the theory. In this case, the multiple-extrema form of the effective potential holds for almost all of the black holes emerging in the theory. As expected, these extremal points (wells and peaks) are translated into a similar nonmonotonic behavior of the GB factors and to their overall suppression, especially as we reach the largest values of $\alpha/M^2$.

The study of the QNMs was also performed in this case but only in the full-theory approach in order to derive the most accurate results. As in the previous theory, we confirmed that the multiple-extrema form of the potential affects also the QNM curve, in accordance with previous studies \cite{Antoniou:2024gdf}. One could safely conclude that black holes in this theory with the highest hair-sourcing parameter are characterized by a larger real part and a smaller (less negative) imaginary part of their quasinormal frequency, thus having a larger damping time, compared to the solutions that lie closer to the GR limit. 

Based on our analysis, we may thus conclude that scalarized black-hole solutions may exhibit distinct features compared to their GR analogs such as modified (suppressed or nonmonotonic) greybody curves, absorption cross sections with distinctly different low and high-energy limits and altered quasinormal frequencies with respect both to the oscillating frequencies and damping times. These features are therefore worth studying, as they may provide clear signatures for the emergence of scalarized black-hole solutions and for the existence itself of a more fundamental theory of gravity.

\acknowledgements
G.A. acknowledges support from the INFN TEONGRAV initiative.~T.D.P. acknowledges the support of the Research Centre for Theoretical Physics and Astrophysics at the Institute of Physics, Silesian University in Opava.

%\appendix

\begin{center}
\small\textbf{DATA AVAILABILITY}
\end{center}

The data that support the findings of this article are not publicly available. The data are available from the authors upon reasonable request.

\bibliography{bibnote}
\end{document}